\documentclass[aps,prb,twocolumn,longbibliography,floatfix]{revtex4-2}

\usepackage{multirow}
\usepackage{siunitx}
\usepackage{lmodern} 
\usepackage{mathtools}

\usepackage{chngcntr}

\usepackage[bookmarks, colorlinks=false]{hyperref}
\hypersetup{colorlinks=true, allcolors=blue}

\usepackage{bbold}
\usepackage{bm}
\usepackage{upgreek}

\usepackage{xurl} 

\begin{document}

\renewcommand{\vec}[1]{\boldsymbol{\mathrm{#1}}}

\renewcommand{\tensor}[1]{\boldsymbol{\mathrm{#1}}}

\title{Atomistic simulations of irradiation damage on the engineering timescale: Examining the dose rate effect in tungsten}
\author{Max Boleininger}
\email{max.boleininger@ukaea.uk}
\affiliation{United Kingdom Atomic Energy Authority, Culham Campus, Abingdon, Oxfordshire, OX14 3DB, United Kingdom}
\author{Daniel R. Mason}
\affiliation{United Kingdom Atomic Energy Authority, Culham Campus, Abingdon, Oxfordshire, OX14 3DB, United Kingdom}
\author{Thomas Schwarz-Selinger}
\affiliation{Max Planck Institute for Plasma Physics, Boltzmannstr.~2, 85748, Garching, Germany}
\author{Pui-Wai Ma}
\affiliation{United Kingdom Atomic Energy Authority, Culham Campus, Abingdon, Oxfordshire, OX14 3DB, United Kingdom}

\begin{abstract}
The change in materials properties subjected to irradiation by highly energetic particles strongly depends on the irradiation dose rate. Atomistic simulations can in principle be used to predict microstructural evolution where experimental data is sparse or unavailable, however, fundamental limitations of the method make it infeasible to replicate the experimental timescale spanning from seconds to hours. Here, we present an atomistic simulation method where the motion of vacancies is accelerated, while the fast degrees of freedom are propagated with standard molecular dynamics. The resulting method is free of adjustable parameters and can predict microstructural evolution under irradiation at elevated temperatures. Simulating the microstructural evolution of tungsten under irradiation at dose rates of $10^{-5}$, $10^{-4}$, and {$10^{-3}$\,dpa/second}, we find that increasing the temperature or reducing the dose rate primarily results in a reduction of the steady-state defect concentration, in qualitative agreement with deuterium retention and post-irradiation resistivity recovery experiments. The formation of a nanoscale void is observed if a system initially containing a large dislocation loop is irradiated. We present a minimally simple rate theory model which reproduces the time-dependent defect concentration and volume swelling behaviour obtained from the simulations.
\end{abstract}

\maketitle

\section{Introduction}
Radiation damage is introduced to atomic systems following a simple schema; when high-energy particles collide with atoms of the material, they may impart sufficient kinetic energy to displace them from their lattice site. If sufficient energy is transferred to the recoil atom, this leads to a self-similar cascade of secondary, tertiary, and so on, atomic collisions. Introducing the generation of radiation damage into atomistic simulations simply involves assigning individual atoms a large amount of kinetic energy and propagating the resultant trajectories in time \cite{granberg2016mechanism}. The various phenomena observed in the context of radiation damage simulations, such as the distinct phases of the collision cascade \cite{Calder2010}, the saturation of radiation damage at high dose \cite{boleininger2023microstructure}, and the rapid relaxation of external stresses \cite{feichtmayer2024fast} emerge from the collective behaviour of atoms over the course of propagation. Using this approach, the recent years saw the generation of irradiation microstructures in qualitative and quantitative agreement with experimental observations of materials properties such as thermal resistivity \cite{mason2021parameter}, lattice strains \cite{Mason2020}, vacancy concentration \cite{boleininger2023microstructure}, and low-temperature irradiation creep \cite{feichtmayer2024fast}. These were genuine predictions, in the sense that no adjustable parameters or fitting were required.

Some external input may be required, for example, \textsc{spectra-pka} \cite{gilbert2015energy} can be used to generate a recoil spectrum so that atoms can be assigned recoil energies consistent with ion, electron, or neutron irradiation, while \textsc{srim} \cite{ziegler2004srim} can be used to generate input for the electronic stopping model. These inputs affect the quantitative outcome of the simulations. The most critical external input is the choice of the model for the interatomic forces. In large-scale atomistic simulations, these forces are typically described by empirical potentials, parameterised such that materials and defect properties are as close as possible to density-functional theory. Recent years also saw the emergence of machine-learned interatomic potentials, which can offer forces with density-functional theory accuracy at a fraction of the computing time \cite{goryaeva2021efficient}. 

However, what gives atomistic simulations such predictive power---the explicit propagation of atomic trajectories---simultaneously leaves the method incapable of simulating the experimentally relevant timescales of seconds, minutes, days, and longer. Atoms in a metal vibrate at a frequency of around $10^{13}\,\mathrm{Hz}$, and so to resolve this oscillation, a propagation time-step order a hundredth of the oscillation period is needed. With contemporary high-performance computing, it is possible to propagate atomic trajectories on the order of a few billion steps, but even that results in a total simulated time of only a few microseconds; this is still far removed from the timescale of a tensile test, or the service lifetime of a fusion reactor component. Under certain conditions, such as high dose rates or low irradiation temperatures, this presents no concern as microstructural evolution is primarily driven by the irradiation damage generation process, which is accessible on the simulated time-scale \cite{boleininger2023microstructure}. However, simulations at elevated irradiation temperatures, where a significant amount of thermal relaxation is expected to occur, have conventionally been out of reach. 

Coarse-grained methods, such as rate theory \cite{Glowinski_JNM_1973,Mansur_JNM_1978,Brailsford_JNM_1972,Westmoreland_RE_1975,Xu_CMS_2016,Li_JAC_2021,Saidi_JNM_2021} or object kinetic Monte Carlo \cite{Xu_JCAMD_1999,Chiapetto_NME_2016,Li_JNM_2022,Domain_JNM2004,MartinBragado_CPC2013,Castin_JNM2017,SivakCR2010,SivakJNM2011,CaturlaCMS2019,JimenezCMS2016,MockJNM2019} have been employed to simulate the time-evolution of defect concentrations under various conditions. In these methods, the evolution of selected defect entities, such as point defects, dislocation loops, and voids, are simulated following a predefined set of rules. Although these coarse-grained methods can simulate microstructural evolution spanning the timescale from microseconds to hours, they require prior knowledge of the defect types emerging during evolution and an appropriate selection of input parameters describing properties of these defects. While quantitive predictions derived from these methods are highly sensitive to the choice of the aforementioned defect types and properties, they offer valuable insight into whether phenomena observed in experiments are consistent with simple dynamic models. 

Here, we present a hybrid atomistic/coarse-grained method where faster degrees of freedom are handled by explicit time-propagation of atomic trajectories while the motion of vacancies is accelerated using a kinetic Monte Carlo scheme. The rationale behind this approach is that microstructural evolution of irradiated materials exhibits a separation of time-scales \cite{keys1968high}, with interstitial-type defects diffusing many orders of magnitude faster than vacancies. While the fast degrees of freedom are evolved in a standard atomistic simulation, we intermittently accelerate the next rate-limiting mode of microstructural evolution, which is vacancy migration. The resulting method can be used to predict microstructural evolution under irradiation at moderately elevated temperatures, where thermal diffusion of vacancies is sufficiently active to lead to thermal recombination of defects and formation of void clusters. While the method relies on specific assumptions on the timescales governing microstructural evolution, it is free of adjustable parameters. We apply the method to quantify the influence of temperature on the irradiation microstructure in tungsten at dose rates of $10^{-5}$, $10^{-4}$, and {$10^{-3}$\,dpa/s}. The data is matched to a minimally simple rate theory model, allowing to predict the vacancy concentration and volume swelling as a function of temperature and dose rate valid up to doses of order {1\,dpa}, before the occurrence of breakaway swelling. Current limitations of the method are also described. 

We conclude with a comparison to experiment, which shows a qualitatively correct reduction in defect content at elevated temperature, with simulation slightly overestimating the degree of defect recombination. We attribute this to a lack of impurities in our model.

\section{Theory}

\subsection*{Rationale of the model}

\begin{figure*}[t]
\includegraphics[width=0.9\textwidth]{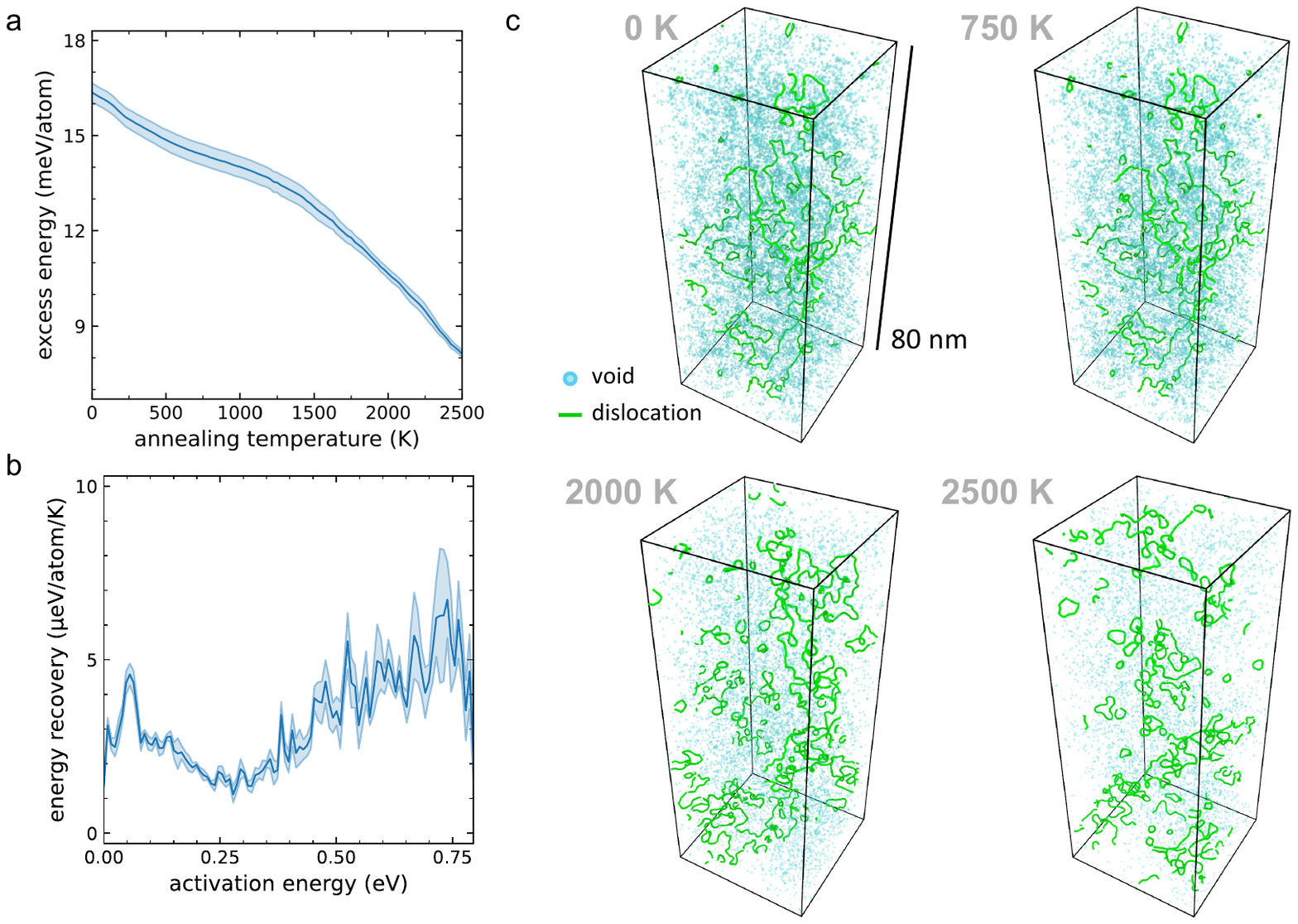}
\caption{Recovery of irradiated tungsten through rapid annealing. \textbf{a} The microstructure of a tungsten crystal, here irradiated to 0.32 dpa, stores potential energy. As the crystal is heated, the microstructure evolves and coarsens, releasing the excess energy, shown here on top of the thermal energy. \textbf{b} The rate of energy release (bottom) allows inference on the activation energy by which the recovery proceeds. The simulated heating rate is unrealistically high with {$2.5\times 10^{11}$\,K/s}; in experiment, recovery is expected to occur at lower temperature. Shown here is the mean recovery over five simulations, each starting with a different {0.32\,dpa} microstructure. Shaded areas indicate the standard error. \textbf{c} Microstructural view of tungsten during post-irradiation annealing. Pictured are dislocation lines (green: $\frac{1}{2}\left\langle 111 \right\rangle$) and void-type defects (blue). As monovacancies become mobile at around {1500\,K} at this heating rate, stage-III recovery commences: vacancies coalesce into vacancy-type dislocation loops and nanoscale voids, clearing up the microstructure. Shaded areas indicate the standard error over 5 independent annealing simulations.}
    \label{fig:1}
\end{figure*}

The accelerated simulation method is built on the assumption that irradiated microstructures evolve on multiple starkly different timescales. Consider the following illustrative simulation: An irradiated single-crystal tungsten, simulated by applying successive recoils up to {0.32\,dpa} ("displacements per atom") at effectively cryogenic conditions, see Ref.~\cite{feichtmayer2024fast}, is heated up. As the crystal is annealed, the irradiation microstructure coarsens and recovers, releasing its stored potential energy. The anneal begins with a temperature at absolute zero and proceeds with a rapid heating rate of {$2.5 \times 10^{11}$\,K/second} until a temperature of {2500\,K} is reached. In Fig.~\ref{fig:1} we plot the stored energy per atom relative to a perfect tungsten crystal at identical temperature over the course of the anneal, as well as the energy release rate over the anneal, with the temperature scale converted into an effective activation energy $E_\mathrm{A}$ via Arrhenius’ relation:
\begin{equation}
E_\mathrm{A} = -k_\mathrm{B} T \ln\left(\nu_\mathrm{a}/\nu_\mathrm{D}\right),
\end{equation}
where $\nu_\mathrm{D} = 10^{13}\,\mathrm{Hz}$ is the Debye frequency and $\nu_\mathrm{a} = 2.5 \times 10^{11}\,\mathrm{K/second}$ is the characteristic timescale of the heating rate.

While some recovery occurs at a low activation energy peaked at {0.05\,eV}, most of it occurs at a much higher activation energy peaked at {0.75\,eV}. This activation energy can be rationalised by considering that for this tungsten interatomic potential \cite{mason2017empirical} the monovacancy migration barrier is {1.53\,eV}, \textit{c.f.} {1.73\,eV} for density-functional theory (DFT) \cite{Ma_PRM_2019_vacancy}, and that the recovery of monovacancy defects in the first instance involves the recombination between vacancy and self-interstitial defects. This is a 2nd order reaction kinetics process, occurring with an effective activation energy of half the vacancy migration energy which matches the observation here. This relationship between higher-order defect kinetics and activation energies is well known in the thermal desorption spectroscopy and post-irradiation resistivity recovery communities \cite{anand1978recovery}. It is also known that for a lower heating rate the recovery peak becomes narrower and shifts towards a lower temperature. In resistivity recovery experiments, this peak is known as stage-III recovery, typically occurring at a temperature of around {200\,$^\circ$C} to {300\,$^\circ$C} for tungsten, depending on the heating rate (\textit{c.f}. {2000\,$^\circ$C} in the simulation due to the the rapid annealing rate used here) \cite{keys1968high, reza2022thermal}.

Given that there is quite little recovery between the two identified recovery stages, we do not necessarily need to accelerate the time propagation of the microstructure for all thermally activated processes equally. At elevated temperature, for example {300\,$^\circ$C} in tungsten, a monovacancy is expected to readily hop from lattice to lattice site with a rate of about {0.01\,Hz}. Over the timescale of hours, the vacancy degrees of freedom are expected to undergo significant evolution at this temperature. Degrees of freedom with a very low activation energy, for example {0.05\,eV}, would evolve with a comparatively enormous rate of {$10^{12}$\,Hz} at this temperature, and hence any meaningful evolution of these degrees of freedom concludes within a few picoseconds; this timescale is easily accessible in a standard atomistic simulation, and hence does not need acceleration. 

The above discussion extends to most base-centred cubic (BCC) metals. DFT calculations find self-interstitial atom migration energies mostly in the order of meV \cite{Ma_PRM_2019_Universality}, or {10\,meV} for Cr, Mo, and W \cite{Ma_PRM_2019_symmetry_broken}, with notable exception of Fe with {0.34\,eV} \cite{Fu_PRL_2004, Ma_PRM_2019_Universality} due to the effect of magnetism. While the migration energy of vacancies in different materials can vary substantially, for the BCC transition metals it ranges from 0.5 to {2.0\,eV} \cite{Ma_PRM_2019_vacancy}, which confirms the general notion that self-interstitial atoms diffuse thermally at much lower homologous temperature than vacancies. This separation of timescales appears to be a general feature of BCC metals. While there may be processes active within the time-scale gap, these cannot be simply addressed within this single split picture. We present a method that is suited for describing the recovery of elemental metals with irradiated microstructure, rather than a general time acceleration method.

\subsection*{Acceleration method}

In order to propagate the vacancy system by a simulated time of $t_\mathrm{sim}$, we follow the accelerated atomistic simulation method illustrated in Fig.~\ref{fig:2}. It proceeds as follows: First, the system is propagated with standard molecular dynamics for a short amount of time $\Delta t_\mathrm{MD}$, a few picoseconds, to relax the degrees of freedom with low activation energy. Next, the system is analysed for monovacancies (as divacancies are not binding in tungsten \cite{heinola2017stability}, they are treated as two neighbouring monovacancies). Finally, the vacancy system is propagated by a time $\Delta t_\mathrm{vac}$ by means of the rejection-sampling kinetic Monte Carlo (rkMC) algorithm  \cite{voter2007introduction}, with vacancy hopping rates depending on the local stress to account for stress-biased diffusion. The vacancy propagation time is chosen such that the rejection rate averages {20\,\%}, as $\Delta t_\mathrm{vac} \sim (5Z\nu)^{-1}$, where
\begin{equation}\label{eq:arrhenius}
\nu = \nu_\mathrm{D} \exp\left(-\frac{E_\mathrm{m}}{k_\mathrm{B} T}\right)
\end{equation}
is the hopping rate of a vacancy to one of its adjacent sites ($E_\mathrm{m}$\,=\,1.73\,eV, as estimated by Arrhenius' law, and $Z=8$ is the number of vacancy hopping sites in a BCC crystal. These three steps are repeated for $N_\mathrm{acc}$ cycles until the total simulated time $t_\mathrm{sim} = N_\mathrm{acc} \Delta t_\mathrm{vac}$ is reached. The rkMC algorithm is presented in more detail in the following section.

\subsection*{Implementation}

The first step is the detection of vacancies in the microstructure. We note that it is crucial that the algorithm never misidentifies a larger vacancy cluster, \textit{i.e.} size trivacancy or larger, as a mono or divacancy. Since divacancies in tungsten are not strongly bound, the nucleation of voids requires a rare coincidental meeting between three vacancies to form a trivacancy. If such a larger cluster is then misidentified as one or more monovacancies, the acceleration algorithm could dissociate the cluster by driving the vacancies apart again, thereby irrevocably disturbing the delicate clustering process. In addition, the vacancy detection algorithm needs to conclude within a runtime comparable to, or faster than, the runtime of the MD propagation step $\Delta t_\mathrm{MD}$ in order to keep the overhead of the algorithm small. 

\begin{figure}[h]
\centering
\includegraphics[width=.9\columnwidth]{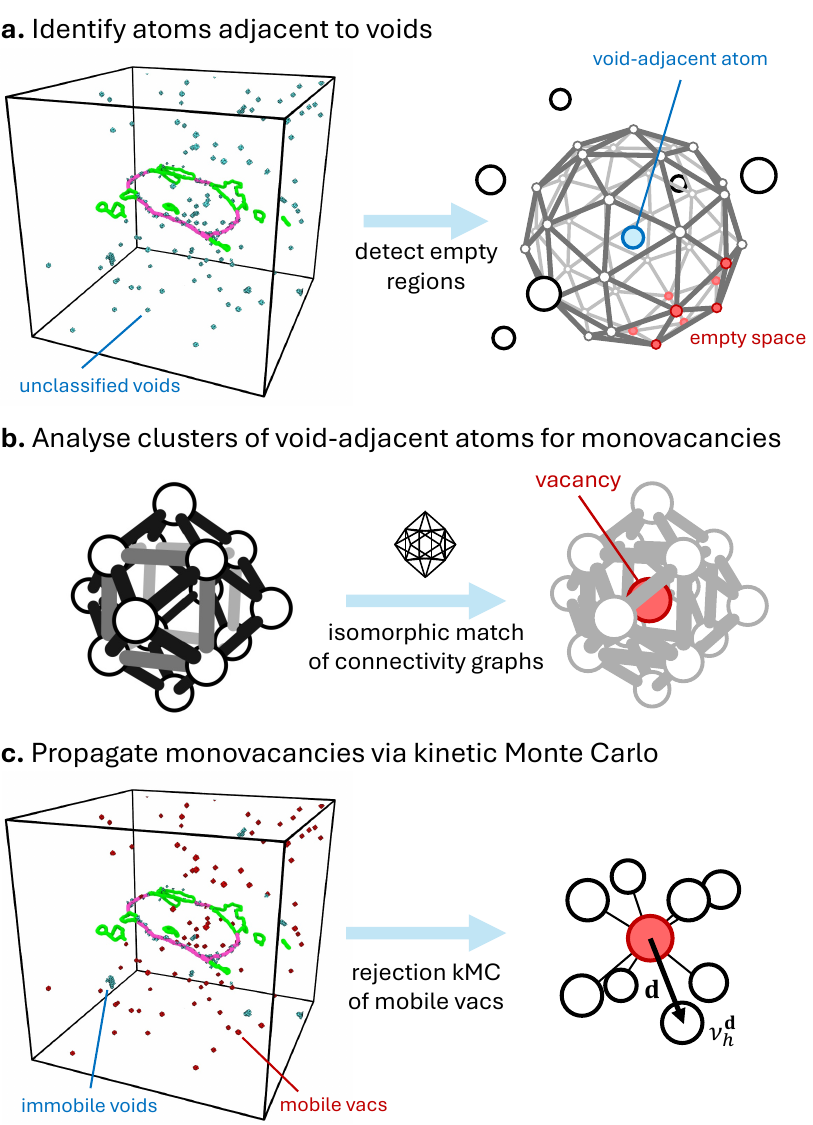}
\caption{Vacancy-acceleration algorithm. \textbf{a} Left: from the microstructure, shown are voids and dislocation lines,
we identify atoms adjacent to void regions. Right: an atom is classified as adjacent to a void if any point (red) in a radius of $|\frac{1}{2}\left\langle 111 \right\rangle|$ from the atom is further than $0.75 \times |\frac{1}{2}\left\langle 111 \right\rangle|$ from any other atom. \textbf{b} Atoms adjacent to voids are clustered by proximity. For each cluster, a connectivity graph is constructed. If the graph is isomorphic to the precomputed graphs of the monovacancy or the two unique divacancy configurations, then the corresponding defect is identified. \textbf{c} As only monovacancies and divacancies are accelerated here, larger void structures are effectively immobile. For each vacancy, the hopping rates $\nu_{h}^{\vec{d}}$ to the positions of its first neighbours $\vec{d}$ are computed, accounting for the virial atomic stress at its current and target positions. The outcome of the random vacancy migration event is evaluated with a rejection kinetic Monte Carlo algorithm. Vacancy migration is considered a stochastically independent and spatially local process.}
    \label{fig:2}
\end{figure}

The void detection algorithm is based on reusing the (full) neighbour list from the MD simulation to detect empty space around each atom. The algorithm is outlined in Fig.~\ref{fig:2}. If an atom is adjacent to a void in a BCC metal, then there should be free space in the crystal around the first neighbour distance of $\sqrt{3}/2 a$ to the atom, where {$a = 3.165$\,\AA} is the tungsten lattice constant. In practice, we sample points on a geodesic sphere of radius $\sqrt{3}/2 a$ centred on each atom. If a point on the geodesic is further than $0.75\times \sqrt{3}/2 a$ to any other atom in the neighbour list of the central atom, then the central atom is adjacent to a void. Next, all void-adjacent atoms are clustered, and connectivity graphs for each cluster are built with a cut-off of $1.1 a$. The cut-off is chosen as larger than $a$ to improve the accuracy when thermal fluctuations are present. Next, the connectivity graphs are compared to precomputed connectivity graphs of the monovacancy and the two unique divacancy structures. If an isomorphic match is detected, then the cluster of void-adjacent atoms is surrounding a monovacancy or a divacancy. In the case of a monovacancy, the monovacancy coordinate is placed in the centre of the cluster, and in the case of a divacancy, the vacancies are placed along the long axis determined by the second central moments of the cluster atom coordinates. This method can be applied to detect larger vacancy clusters by including more precomputed connectivity graphs---in practice, we found that enumeration of the unique number of vacancy cluster configurations up to size 5 was feasible. The specificity and detection accuracy is benchmarked in Appendix~\ref{app:vacdetection}.

Once the vacancies are detected, the next step is to propagate their positions with the rkMC algorithm. From the previous vacancy clustering step, we already have the list of potential hopping sites available. We now evaluate the list of hopping rates of each individual vacancy to their respective hopping sites. To account for the effect of elastic stress on the vacancy hopping rate, consider the hopping rate of a vacancy at position $\vec{r}_0$ to a neighbouring site $\vec{r}_0 + \vec{d}$. The hopping rate is estimated by the Kang–Weinberg model \cite{kang1989dynamic, mason2019atomistic}
\begin{equation}\label{eq:hoppingrate}
\nu_h^{\vec{d}} = \nu_\mathrm{D} \exp\left(
    -\tfrac{
E_\mathrm{m} 
+ \mathrm{tr}\left( 
    \tensor{\upsigma}_0 \cdot \tensor{\Omega}_0
    - \frac{1}{2}(
    \tensor{\upsigma}_{\vec{0}} + \tensor{\upsigma}_{\vec{d}})
    \cdot \tensor{\Omega}_\mathrm{sd}^{\vec{d}} 
\right)
}
{k_\mathrm{B} T}
\right),
\end{equation}
where $\tensor{\upsigma}_0$ and $\tensor{\upsigma}_{\vec{d}}$ are the virial atomic stress tensors at the vacancy and hopping site positions, respectively, and $\tensor{\Omega}_0$ and $\tensor{\Omega}_{\mathrm{sd}}^{\vec{d}}$ are the relaxation volume tensors of the vacancy at the ground state and saddle point, respectively. Here, the stress tensor at the saddle point is approximated by the average of the stress tensors at the vacancy and hopping sites. The tensor $\tensor{\Omega}_0$ is known from DFT \cite{Ma_PRM_2019_vacancy}:
\begin{equation}
\tensor{\Omega}_0 = \Omega_{\mathrm{iso}}^\mathrm{(eq)} \tensor{I}_3 
\end{equation}
where $\tensor{\mathrm{I}}_3$ is the identity matrix of size 3 and  $\Omega_{\mathrm{iso}}^\mathrm{(eq)} = -0.115$ atomic volumes. The general expression for the relaxation tensor at the saddle-point for a hop in an arbitrary direction is:
\begin{equation}
\begin{aligned}
\tensor{\Omega}^{\vec{d}}_{\mathrm{sd}} 
&= \left(\Omega_{\mathrm{iso}}^\mathrm{(sd)} + 2 \Omega_{\mathrm{aniso}}^\mathrm{(sd)} \right) 
\left(
    \hat{\vec{d}}\otimes \hat{\vec{d}}
\right) \\
&\phantom{=}+ \left(\Omega_{\mathrm{iso}}^\mathrm{(sd)} -  \Omega_{\mathrm{aniso}}^\mathrm{(sd)} \right) 
\left( 
    \hat{\vec{u}}\otimes \hat{\vec{u}}
+   \hat{\vec{v}}\otimes \hat{\vec{v}}
\right)
\end{aligned}
\end{equation}
where $\Omega_{\mathrm{iso}}^\mathrm{(sd)} = -0.099$ and $\Omega_{\mathrm{aniso}}^\mathrm{(sd)} = -0.094$ in atomic volumes, and vectors $\vec{u}$ and $\vec{v}$ are given by 
\begin{equation}
\begin{aligned}
\vec{u} &= \left(-\frac{d_y + d_x}{d_z}, 1, 1 \right) \\
\vec{v} &= \left(
    1, 
    \frac{d_x^2 + d_z^2 + d_y d_z}{d_x (d_z - d_y)}, 
    -\frac{d_x^2 + d_y^2 - d_y d_z}{d_x (d_z - d_y)}
    \right).
\end{aligned}
\end{equation}
The above expression is derived by rotating the relaxation volume tensor for a hop in $[111]$ direction to an arbitrary direction $\vec{d}$. This formulation does not require prior identification of the crystal orientation, and can hence be applied to polycrystalline systems. To avoid numerical issues in evaluating $\vec{v}$ when $d_y = d_z$, we note that the singularity is removable in $\hat{\vec{v}}$:
\begin{equation}
\lim_{d_y\rightarrow d_z^\pm} \hat{\vec{v}} = \left(
    0, 
    \mp\tfrac{d_x^2 + 2 d_z^2}{\sqrt{2 d_x^4 + 4 d_x^2 d_z^2 + 4 d_z^4}}, 
    \pm\tfrac{d_x^2}{\sqrt{2 d_x^4 + 4 d_x^2 d_z^2 + 4 d_z^4}}
    \right),
\end{equation}
and that the tensor product $\hat{\vec{v}}\otimes \hat{\vec{v}}$ is invariant to the direction of the limit. The relaxation volume tensor is therefore well defined for any hopping direction.

We note that Eq.~\eqref{eq:hoppingrate} for the vacancy hopping rate does not include any quasiharmonic or anharmonic effects. To account for these, we require first-principles calculations of the temperature-dependent migration barrier and relaxation volume tensors. Recent progress in simulating the anharmonic vacancy migration in tungsten with a machine-learned potential indicates that such data may be available in the near future \cite{zhang2025ab}.

Before each rkMC step, we extract the coordinates $\{\vec{r}_i\}$ and virial atomic stresses $\{\tensor{\sigma}_i\}$ of all atoms within a range of vacancies, from which we obtain the virial stress at an arbitrary position $\vec{R}$, which may for example be the vacancy or hopping site positions, by kernel density estimation:
\begin{equation}\label{eq:stress_field}
    \tensor{\upsigma}_{\vec{R}} =
\frac{\sum_i f(|\vec{R}-\vec{r}_i|) \tensor{\upsigma}_{i}}
    {\sum_j f(|\vec{R}-\vec{r}_j|)},
\end{equation}
where we chose the kernel function
\begin{equation}
    f(r)=\frac{1}{2}
        \left(
            \sin\left(
                \dfrac{\pi r}{r_0}+\dfrac{\pi}{2}
                \right) + 1 
        \right) \Theta(r_0 - r),\label{eq:KDE_fn}
\end{equation}
where $\Theta(x) = 1$ if $x > 0$, otherwise $0$, is the Heaviside function, and $r_0 = 5.8\,\text{\AA}$ is a cutoff radius chosen between the fifth and sixth nearest neighbours. We note that the virial atomic stress is often reported in dimensions of pressure times volume, thereby requiring division by the atomic volume. Here, this is avoided by representing the relaxation volume tensors in units of atomic volume, resulting in the product of virial stress and relaxation volume returning the correct dimensions. The vacancy acceleration method is tested by comparison with standard molecular dynamics for the diffusion of a monovacancy with and without external stress in Appendix~\ref{app:kmc}.

The rkMC iteration is performed after the hopping rates $\{v_h^{\vec{d}}\}$ for all vacancies have been computed. In a given time interval $t_\mathrm{vac}$, the expectation value for the number of hopping events of a given vacancy is $\left\langle n\right\rangle = t_\mathrm{vac} \sum_h v_h^{\vec{d}}$. In a compromise between keeping the number of propagation steps low and reducing the probability of multiple hopping events occurring within the rKMC propagation time step, we choose the time $t_\mathrm{vac} = (5Z\nu)^{-1}$ such that $\left\langle n\right\rangle \approx 0.2$, where the net vacancy hopping rate is estimated by the hopping rate in a stress-free crystal. As we do not account for stress, the actual acceptance rate recorded in our simulations varies between {15\,\%} and {30\,\%}. For each vacancy, we generate a uniform random number $0 <r \leq 1$. If $r \leq 1 - \mathrm{exp}\left(-t_\mathrm{vac}\sum_h v_h^{\vec{d}} \right)$, then the vacancy undergoes a migration event, chosen from the set of possible hops, with probability weighted by the hopping rate. Otherwise, the vacancy remains at its site for this iteration. The migration event is implemented by creating an atom at the position of the vacancy and deleting the atom at the vacancy hopping site.

The method for simulating overlap cascades is described in detail in Ref.~\cite{feichtmayer2024fast}. In order to apply the acceleration method to an irradiation damage simulation, we note that an atomic recoil increments the dose by $\Delta \phi = 0.8 T_\mathrm{d}/(2E_\mathrm{d} N_\mathrm{atoms})$ in units of NRT dpa \cite{norgett1975proposed}, where $T_\mathrm{d}$ is the damage energy evaluated from the recoil energy using the Lindhard model \cite{norgett1975proposed} and $E_\mathrm{d} = 90\,\mathrm{eV}$ the threshold displacement energy for tungsten. Hence, for a given target dose rate of $\dot{\phi}$ in units of {dpa/s}, in between each atomic recoil, the vacancy subsystem needs to be propagated for a total time of $t_\mathrm{sim} = \Delta\phi/\dot{\phi}$, or a total number of $N_\mathrm{acc} = t_\mathrm{sim}/\Delta t_\mathrm{vac}$ rkMC steps.

The MD simulation are performed in \textsc{Lammps} \cite{plimpton1995fast}, which is compiled as a shared library and driven from Python \cite{python3} along with the vacancy detection and rkMC algorithms. 

\section{Results}

We performed irradiation damage simulations to quantify the effect of dose rate on the defect concentration. We simulated the irradiation of an initially perfect, pure tungsten single crystal consisting of 80\,$\times$\,80\,$\times$\,80 body-centred crystal unit cells (25\,$\times$\,25\,$\times$\,25\,nm$^3$), in total containing 1,024,000\,atoms, with periodic boundary conditions applied to all three directions. We simulated the following conditions of dose rate $\dot{\phi}$ and temperature $T$:
\begin{itemize}
\item $\dot{\phi} = 10^{-3}$\,{dpa/s}, $T=$ 600, 650, 700, 750, and {800\,K}
\item $\dot{\phi} = 10^{-4}$\,{dpa/s}, $T=$ 550, 600, 650, 700, and {750\,K}
\item $\dot{\phi} = 10^{-5}$\,{dpa/s}, $T=$ 500, 550, 600, 650, and {700\,K}
\end{itemize}
The temperature ranges were chosen to sample cascade-driven microstructural evolution at lower temperatures, and diffusion-driven evolution at higher temperatures, for a given dose rate. To test for the effect of a sink, we also repeated the higher-temperature {$10^{-4}$\,dpa/s} dose rate simulations with a tungsten crystal initially containing a interstitial-type dislocation loop of {13\,nm} diameter with $\left\langle 100 \right\rangle$ Burgers vector, corresponding to a dislocation density of {2.5\,$\times$\,$10^{15}$\,$/\mathrm{m}^{2}$}.

\begin{figure*}[t]
\centering
\includegraphics[width=0.9\textwidth]{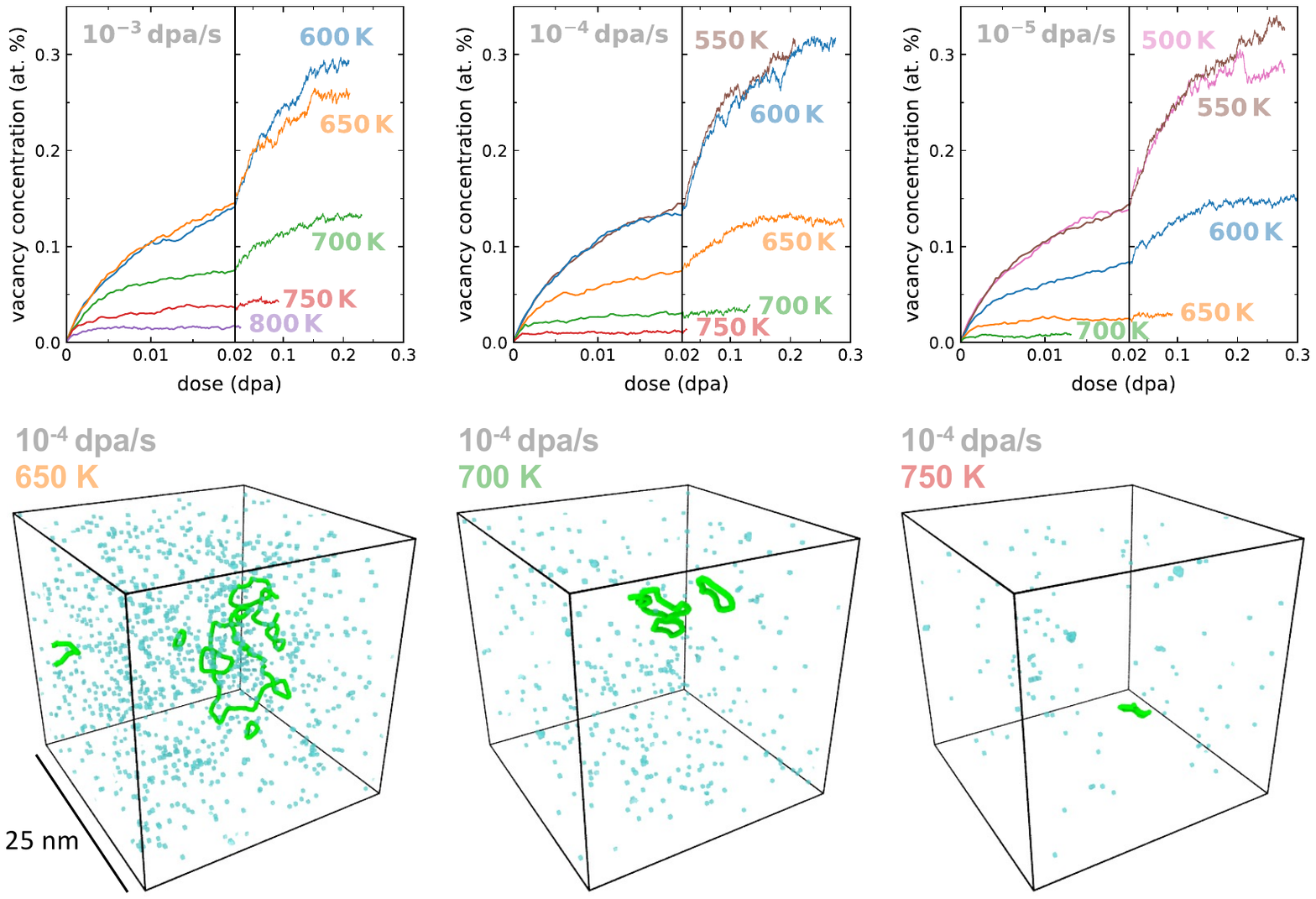}
\caption{Simulated irradiation of tungsten as a function of dose, dose rate and temperature. Top row: The vacancy concentration is found to decrease with increasing temperature and decreasing dose rate. Note the broken x-axis scale. Bottom row: Simulation snapshots of the highest simulated doses for the {$10^{-4}$\,dpa/s} dose rate, 0.30, 0.13 and {0.03\,dpa}, at a temperature of 650, 700, and {750\,K}, respectively. Pictured are dislocation lines (green: $\frac{1}{2}\left\langle 111 \right\rangle$), here all of interstitial type, and void-type defects (blue). Some void clustering is observed at {750\,K}.}
    \label{fig:3}
\end{figure*}

In all simulations, a mono-energetic recoil spectrum with {10\,keV} was used, resulting in a dose increment of {$3.5\times 10^{-5}$\,dpa} per recoil. A barostat was applied to keep the system-averaged stress to zero, using the approach by Shinoda \textit{et al.} \cite{shinoda2004rapid} implemented in \textsc{Lammps}. The system was propagated for {10\,ps} after each atomic recoil to ensure that the collision cascade would conclude. Next, the fast degrees of freedom were relaxed for {1\,ps} in between each of the $N_\mathrm{acc}$ rkMC steps. This process was repeated until the simulations reached a maximum dose of {0.3\,dpa}. The higher temperature simulations however were unable to reach this dose; at high temperature, the vacancies undergo thousands of migration events between each cascade, slowing down the simulation substantially.

\begin{figure*}[t]
\includegraphics[width=0.9\textwidth]{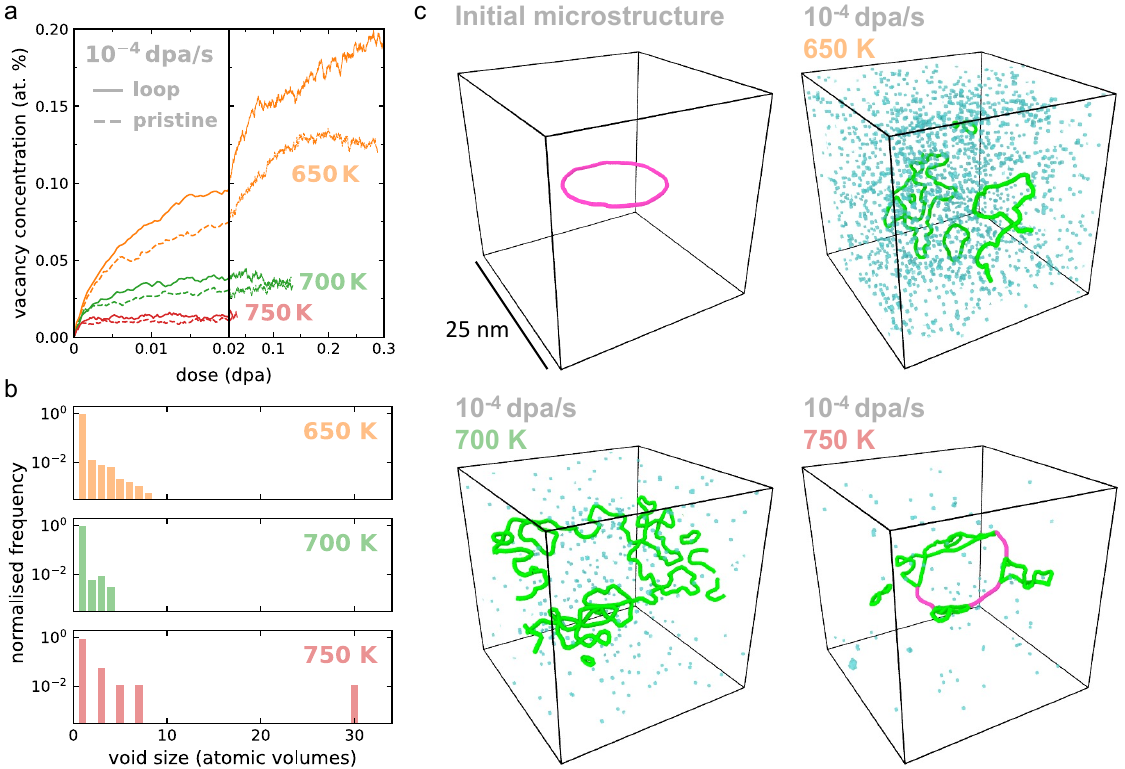}
\caption{Simulated irradiation of tungsten containing an initial $\left\langle 100 \right\rangle$ loop of {13\,nm} diameter as a function of dose and temperature. \textbf{a} The initial dislocation loop appears to act as a sink, absorbing interstitials preferentially, which suppresses Frenkel pair recombination and consequently leads to a higher vacancy concentration. \textbf{b} No void formation is observed in the {650\,K} and {700\,K} structures, while a void of diameter {1.3\,nm} forms in the {750\,K} structure. \textbf{c} Simulation snapshots of the initial microstructure and highest simulated doses at 0.30, 0.14 and 0.03 dpa, at a temperature of 650, 700, and {750\,K}, respectively. Pictured are dislocation lines (green: $\frac{1}{2}\left\langle 111 \right\rangle$, purple: $\left\langle 100 \right\rangle$) and void-type defects (blue). We find that the $\left\langle 100 \right\rangle$ loop is transformed to a $\frac{1}{2}\left\langle 111 \right\rangle$ loop as interstitials are absorbed.}
    \label{fig:4}
\end{figure*}

In Fig.~\ref{fig:3}, we plot the vacancy concentration as a function of dose and temperature for the three dose rates considered here, and show the resultant microstructures at the highest simulated doses for the dose rate of {$10^{-4}$\,dpa/s}. We define `vacancy concentration` here as the void volume fraction expressed in units of atomic percentage, or at.\,\%. The void content is quantified using the method developed by Mason \textit{et al.} \cite{mason2021parameter}. Note that the x-axis is broken into two different linear scales to aid resolving the graphs at low dose. In Fig.~\ref{fig:4}, we compare the vacancy concentration with and without the initial $\left\langle 100 \right\rangle$ dislocation loop for the dose rate of {$10^{-4}$\,dpa/s} and show the final microstructures. We make the following observations: At low temperature or high dose rate, the vacancy concentration saturates to a value of around {0.31\,at.\,\%}. At athermal conditions, when vacancies are effectively immobile in between cascade impacts, the vacancy concentration is driven to a recoil energy dependent steady-state \cite{mason2021parameter, boleininger2023microstructure}. The saturation concentration of {0.31\,at.\,\%} observed here is lower than that found in athermal cascade simulations for the same tungsten potential, \textit{c.f.} {0.34\,at.\,\%} \cite{boleininger2022volume}
and {0.40\,at.\,\%}  \cite{boleininger2023microstructure}, possibly due to the enhanced mobility of interstitial defects at higher temperature. 

The primary effect of increasing the temperature or decreasing the dose rate is a reduction in the steady-state defect concentration of the material. The steady-state concentration emerges from the balance between the continuous generation of radiation defects by high-energy atomic recoil and the continuous recombination between mobile vacancies and interstitials. The secondary effect is that vacancies cluster together, eventually forming small void clusters that remain stable under irradiation. The final effect is that the presence of the sink, here in the form of a $\left\langle 100 \right\rangle$ loop, leads not only to a higher defect concentration than in an otherwise identical simulation without a sink, but also to the formation of larger, nanoscale voids, see the {750\,K} microstructure in Fig.~\ref{fig:4}. 

From the simulations with an initial dislocation loop, we observe that the loop acts a sink for interstitial-type defects, thereby reducing the chances of vacancy-interstitial recombination, and giving each vacancy more time to find a void cluster to attach to. At the same time, the nature of the sink itself transforms over the course of irradiation, with the dislocation loop Burgers vector changing from $\left\langle 100 \right\rangle$ to $\frac{1}{2}\left\langle 111 \right\rangle$ as interstitial defects are absorbed.

\section{Discussion}

\begin{figure*}[t]
\includegraphics[width=0.9\textwidth]{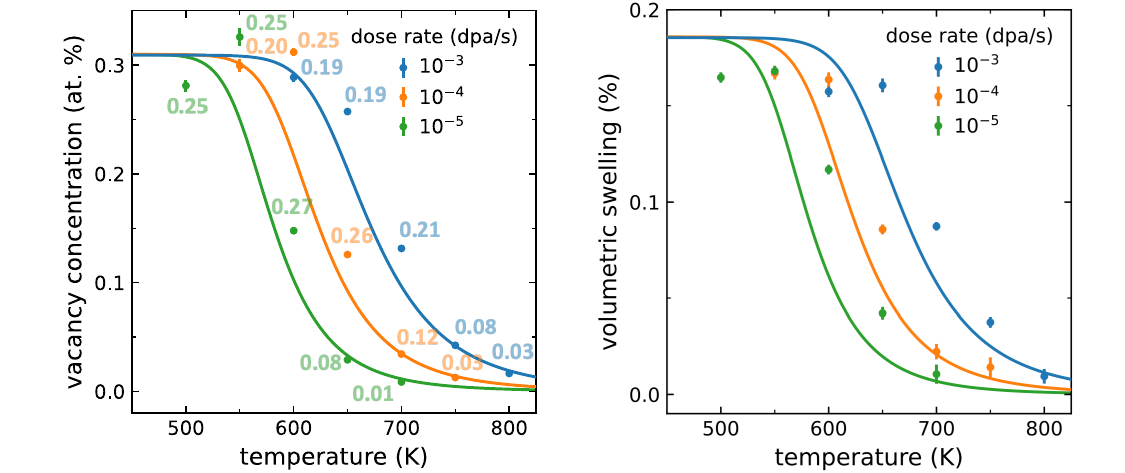}
\caption{Comparing prediction to simulation of irradiation swelling in tungsten. A minimally simple and analytical model of vacancy recombination (line), see text, is consistent with the accelerated irradiation damage simulation (dots), especially under consideration that the model contains no adjustable parameters. The data points and error bars represent the mean and standard deviation of the vacancy or swelling values taken over the latest {20\,\%} of the simulation snapshots, with points labelled by the respective mean dose (dpa). Here, a Debye frequency for tungsten of $\nu_\mathrm{D} = 8 \times 10^{12}\,\mathrm{Hz}$ was used. The thermal expansion strain was removed from the volumetric swelling strain shown here. The labels on each data point indicate the mean dose in dpa. While the doses appear low, the transient component of vacancy evolution appears already concluded, see Fig.~\ref{fig:3}.}
    \label{fig:5}
\end{figure*}

The primary effect, the increased defect recombination, can be rationalised by a minimal rate theory model. Consider first that the highly mobile interstitials generated by cascades cluster together into $\frac{1}{2}\left\langle 111 \right\rangle$-type interstitial loops. As these defects only diffuse one-dimensionally in direction collinear to the Burgers vector, they are unlikely to find many vacancies to recombine with. Vacancies, on the other hand, diffuse three-dimensionally, and are hence more likely to encounter interstitial clusters to recombine with. Assuming that interstitial clusters are effectively sessile in the context of defect recombination, the partial differential equation (PDE) describing time-evolution of the vacancy concentration $c(t)$, which is nominally equal to the Frenkel pair concentration, is written as \footnote[3]{The first term is the 2$^\text{nd}$-order reaction kinetics expression $-4\pi D \eta a R c^2(t)$, where $D= a^2\nu$ is the vacancy diffusion constant and $\eta=2/a^3$ is the bcc atomic density.}:
\begin{equation}
\frac{\mathrm{d}c(t)}{\mathrm{d}t} = -8\pi \nu R c^2(t)
    + \left( 1 - \frac{c(t)}{c_\mathrm{sat}}\right) 
    g \dot{\phi} 
\end{equation}
where $\nu$ is the temperature-dependent hopping rate of a vacancy to one of its neighbouring sites \eqref{eq:arrhenius}, $R$ is the distance at which Frenkel-pairs recombine in units of lattice constants, $g$ is the vacancy generation efficiency in units of {at.\,\%/dpa} and $c_\mathrm{sat}$ is the saturated vacancy concentration in the athermal limit in units of {at.\,\%}. The first term on the right-hand side drives recombination, and was derived assuming that interstitial and vacancy concentrations are equal. The second term is phenomenological, with the purpose of driving the vacancy concentration towards the athermal steady-state, where it saturates to $c_\mathrm{sat}$ \cite{boleininger2023microstructure}. Parameters $\nu$, $R$ are fundamental materials constants, obtainable from first-principles methods such as DFT, while $g$ and $c_\mathrm{sat}$ depend on the material and cascade energy and can either be obtained from overlap cascade simulations or estimated using the arc-dpa model \cite{nordlund2018improving}. For this tungsten potential, with {10\,keV} recoil energy  cascades, we find $g = 0.1125\,\mathrm{at.}\,\%/\mathrm{dpa}$, $c_\mathrm{sat}=0.31\,\mathrm{at.\%}$, and choose $R = 1$. For a constant dose rate, the above PDE has the analytical solution:
\begin{equation}
c(t) = \frac{2 c_\mathrm{sat}}{
    1 + \sqrt{1+\chi}\coth\left(\dfrac{g \dot{\phi} t}{2c_\mathrm{sat}\sqrt{1+\chi}}\right)
    },
\end{equation}
where
\begin{equation}
\chi = 32\pi R 
    \frac{\nu c_\mathrm{sat}^2}{g \dot{\phi}}.
\end{equation}
For large times, the vacancy concentration asymptotically approaches a steady-state:
\begin{equation}
\lim_{t\rightarrow \infty} c(t) = \frac{2 c_\mathrm{sat}}{1 + \sqrt{1 + \chi}}.
\end{equation}

In Fig.~\ref{fig:5}, we show a plot comparing the vacancy concentrations from the accelerated atomistic simulation and the rate theory model. The model is quantitatively consistent with simulation, predicting the extent by which the defect concentration is reduced as vacancies become more mobile at higher temperatures or lower dose rates. The simulation results appear to be shifted by about {20\,K} relative to the model at higher vacancy concentrations. This is expected, as the model is formulated under the assumption that the concentration of interstitials available for recombination is equal to the vacancy concentration. Vacancies can however only recombine with interstitials that lie at the perimeter of defect clusters or dislocation loops, and therefore the recombination rate is overestimated especially for higher defect densities, where large interstitial-type dislocation networks are able to form.

Further, we note that the model can also be used as a minimally simple description of irradiation swelling by converting the defect concentrations into volumetric swelling $\varepsilon_\mathrm{v}$ via the relaxation volumes of the defects. Assuming that interstitials cluster together or join larger pre-existing defect structures, where their relaxation volume asymptotically approaches one atomic volume as the structure grows, and that vacancies appear in the form of either mono-vacancies or small voids, where their relaxation volume is approximately $-0.4$ atomic volumes \cite{mason2019relaxation}, we find for the volumetric swelling strain:
\begin{equation}
\varepsilon_\mathrm{v} \approx (1-0.4) c(t) = 0.6 c(t)
\end{equation}
The swelling model is compared to the volumetric swelling directly measured on the MD simulation box, see  Fig.~\ref{fig:5}. The thermal expansion strain has been removed from the atomistic data for this comparison.

We note that the above model predicts a saturation in vacancy concentration and swelling. Void swelling, that is, a measure for swelling obtained by counting nano-sized and larger voids under a transmission electron microscopy (TEM), is not a feature of the model because it is not accounted for in its formulation; the simulations have not progressed to a dose high enough to accurately quantify this phenomenon. Therefore, the model is expected to offer a useful description of volumetric swelling only up to intermediate dose of order {1\,dpa}.

The secondary effect, the occurrence of void clustering, is in broad agreement with TEM characterisation of tungsten self-ion irradiated at comparable temperature and identical dose rate \cite{hu2021effect}. Void nucleation requires temperatures sufficient for monovacancy diffusion to occur, and hence nanometre-sized voids do not form in standard athermal collision cascade simulations. Similarly, they are not observed in tungsten self-ion irradiated at room temperature under the TEM. The number concentration of voids appears lower than in experiment, however some degree of underestimation is expected as the simulations here do not contain any impurities; tungsten vacancies bind strongly to impurity elements such as C, N, and O, and could thereby catalyse the void nucleation process \cite{song2024bridging}.

As for the third effect, a detailed discussion of this phenomenon requires deeper analysis. Coarse-grained models, such as rate theory or coupled-cluster dynamics, introduce the notion of a 'sink density', with sinks acting as entities that have endless capacity to absorb and thereby remove point defects from the system. Relating this concept back to the atomistic picture is essential in obtaining a more realistic and predictive model of microstructural evolution.

We conclude with a comment on the computational performance of this simulation method. While we succeeded in developing a simple model accounting for the effect of irradiation temperature on microstructural evolution, it is still challenging to perform simulations up to high dose at high temperature. In the same vein that standard atomistic simulations cannot reach experimental timescales because the intrinsic frequency of atomic vibration needs to be numerically resolved, in the vacancy-accelerated simulation method, we find that at high temperature, the vacancies become so mobile that in a given time interval, they are required to undergo an enormous amount of individual migration events. To illustrate the point, we list the expected number of vacancy hops in one hour of real time as a function of temperature, as estimated with Arrhenius’ law, see Tab.~\ref{tab:arrheniusrate}. As the hopping rate increases exponentially with temperature, any acceleration method that resolves individual migration events becomes computationally infeasible at high temperature.

\begin{table}[h]
\small
  \caption{\ Rate of migration events for a tungsten vacancy, as estimated with Arrhenius' law ({$E_\mathrm{A}$\,=\,1.73\,eV})}
\label{tab:arrheniusrate}
\begin{tabular*}{0.48\textwidth}{@{\extracolsep{\fill}}ll}
\hline
temperature (K) & vacancy hops per hour \\
\hline
300  & $5 \times 10^{-13}$ \\
500  & $2 \times 10^{-1}$  \\
700  & $2 \times 10^{4}$   \\
900  & $8 \times 10^{6}$   \\
1100 & $5 \times 10^{8}$   \\
\hline
\end{tabular*}
\end{table}

For the acceleration scheme presented  here, each rkMC cycle is accompanied by 5 MD propagation cycles of {1\,ps} duration each (note the factor 5 is due to the {20\,\%} rkMC acceptance rate). To propagate a single simulation up to one hour at {900\,K} would hence be equivalent to running on the order of $2\times 10^{10}$ MD time steps. Assuming no overhead, with current computing capabilities, it would require on the order of {10,000\,GPU-hours} \footnote[4]{{$5.26\times 10^8$\,atom\,step/s} for an EAM potential on a H200 \textsc{nvidia} card, see \url{https://developer.nvidia.com/hpc-application-performance}, accessed 01/17/2025.} to propagate a 1 million atom system to one hour at {900\,K}. For comparison, this represents about {0.2\,\%} of the annual capacity of the GPU partition of the \textsc{pitagora} cluster (27 PFLOPS/s) \cite{pitagora}. While it is feasible to run a few such simulations, in order to reach longer simulation times at higher temperatures, it is necessary to develop a method where vacancy migration is accelerated in a way that does not require resolution of individual migration events.

\section{Comparison to experiments}

For a number of reasons, it is not straightforward to draw quantitative comparisons between experimental measurements and predictions derived from these simulations. First of all, the simulations predict that the majority of voids appear in the form of small vacancy clusters, with exception of a single large void forming at {750\,K} when a sink is present, see Fig.~\ref{fig:4}. There are few microstructural characterisation techniques that are sensitive to small vacancy clusters. One such technique being positron-annihilation spectroscopy (PAS). However, while it is possible to reconstruct vacancy cluster size distributions with PAS, the absolute void volume fraction is not accessible. Instead, we shall focus on experimental studies reporting on quantities that can act as indirect measures of defect concentration. We consider specifically ion-irradiation experiments, as neutron irradiation introduces additional complexity due to transmutation effects \cite{gilbert2011neutron}.

For our first comparison, we consider depth-resolved deuterium concentration measurements in irradiated tungsten \cite{chrominski2019tem, schwarz2017deuterium} as a proxy for vacancy concentration. In these experiments, tungsten was first subjected to self-ion-irradiation at a given temperature, and then subsequently exposed to a deuterium plasma at elevated temperature over multiple days, allowing the surface-implanted deuterium to diffuse through the damaged layer and bind to the irradiation defects. The temperature is elevated to promote deuterium diffusion, but still kept below stage-III recovery to avoid further microstructural evolution. The deuterium concentration can then be resolved as a function of depth using 3-He nuclear reaction analysis.

Experimental data was obtained for recrystallized, poly-crystalline tungsten irradiated at {290\,K} and {800\,K} with {20.3\,MeV} tungsten ions at an average dose rate of {$10^{-5}$\,dpa/s}, following the procedure outlined in Schwarz-Selinger \textit{et al.} \cite{schwarz2023critical}. Here, however, deuterium loading was performed at {370\,K} instead of {450\,K} to maximise deuterium retention. Deuterium retention data is also available for the same two irradiation temperatures for single-crystalline samples with $\left\langle 100\right\rangle$ \cite{markelj2024first} and $\left\langle 111\right\rangle$ \cite{zavavsnik2025microstructural} surface orientations, here irradiated with {10.8\,MeV} tungsten ions at an average dose rate of {2\,$\times$\,$10^{-4}$\,dpa/s}. We do not consider the difference in ion energy to affect the result, as the recoil spectra are nearly identical below the cascade fragmentation energy \cite{wielunska2020deuterium}. The data for the two surface orientations is plotted together, as the retained deuterium concentration does not appear to be significantly affected by the surface orientation. 

In our simulation, we estimate the number of retained deuterium atoms directly from the void surface area $\Sigma$:
\begin{equation}
N_\mathrm{D} = 5 \frac{\Sigma}{\Sigma_\mathrm{v}},
\end{equation}
assuming that deuterium predominantly binds to internal surfaces, with the surface of monovacancy $\Sigma_\mathrm{v}$ trapping up to 5 deuterium atoms at room temperature \cite{heinola2010hydrogen}, and that the introduction of deuterium does not significantly alter the concentration of microstructural defects. The surface areas are computed using the void detection method presented by Mason \textit{et al.} \cite{mason2021estimate}.

\begin{figure}[t]
\centering
\includegraphics[width=.74\columnwidth]{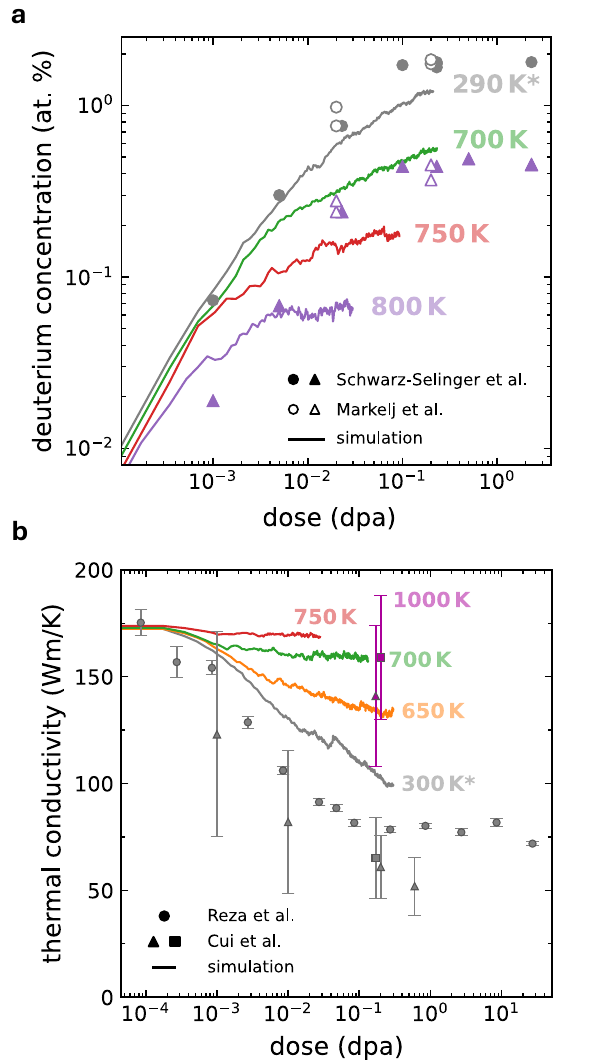}
\caption{Comparison to experiment. \textbf{a} Deuterium retention of tungsten irradiated at elevated temperature. Experimental data is shown for {290\,K} and {800\,K}, in polycrystalline (Schwarz-Selinger \textit{et al.} \cite{schwarz2023critical}) and single-crystalline tungsten (Markelj \textit{et al.} \cite{markelj2024first} \cite{zavavsnik2025microstructural}). At {290\,K}, experiment and simulation are in agreement. At {800\,K}, simulation underestimates the retained amount. \textbf{b} Thermal conductivity of irradiated tungsten. Experimental data is shown for {300\,K} (Reza \textit{et al.} \cite{mason2021estimate}, Cui \textit{et al.} \cite{cui2018thermal}) and for {1000\,K} (Cui \textit{et al.} \cite{cui2018thermal}). As before, experiment and simulation are in broad agreement for {300\,K}. At higher irradiation temperature, simulations underestimate defect concentration, resulting in an overestimation of thermal conductivity. Asterisks indicate that the simulated temperature is higher than stated, but representative for room temperature irradiation, see main text for details.
}
\label{fig:6}
\end{figure}

The measured and simulated deuterium concentrations are shown in Fig.~\ref{fig:6} as a function of dose for different temperatures. To allow for a direct comparison with the {800\,K} experimental irradiation temperature, we use simulation data for a {$10^{-3}$\,dpa/s} dose rate, noting that we would expect more defect recombination in a {800\,K} and {$10^{-4}$\,dpa/s} simulation. As we did not explicitly simulate irradiation at room temperature, we used here the {600\,K} data-set for comparison, labelled as {290\,K} with an asterisk; at this temperature and dose rate, vacancies undergo no thermal diffusion, and therefore the system evolves athermally.

For room temperature irradiation, we find that the predicted deuterium concentration is in good qualitative and quantitative agreement with experiment, which corroborates the findings of a previous study \cite{mason2021parameter}, where a hybrid method for simulating microstructural damage called \textit{cascade annealing} was used, combining cascades and Frenkel pair insertion to reach higher doses. For irradiation at {800\,K}, simulation and experiment both show a reduction in deuterium concentration, however, simulation underestimates the retained amount by at least a factor of three. Another curious point is that similar deuterium retention experiments \cite{schwarz2017deuterium, chrominski2019tem} were performed on tungsten samples irradiated at average dose rates spanning three orders of magnitude, ranging from $5.8\times 10^{-3}$ to $3.9 \times 10^{-6}$\,dpa/s. However, each sample was found to retain a saturated deuterium concentration of about {0.5\,\%} irrespective of dose rate. This effect is not reproduced in our simulations; our predicted steady-state vacancy concentration is dependent on both temperature and dose rate. The notable exception is at low temperature, where the vacancy concentration approaches the athermal steady-state of about {0.3\,\%}. We further note that irradiations were not performed with a continuous beam as recommended by ASTM standards \cite{shao_standardization_2017} but beams were rastered with {1\,kHz} over the sample surface to achieve a laterally homogenous irradiation. As a consequence, the peak damage rate is two orders of magnitude larger than the average rate stated here. For the {290\,K} irradiation we expect rastering not to influence the defect density in this athermal regime but eventually, for a given dose rate at higher temperature and hence higher vacancy mobility, the increased peak dose rate may have an influence on the vacancy density. 

For our next comparison, we consider thermal conductivity measurements of ion-irradiated tungsten. The nanoscale defects generated by irradiation act as electron scattering sites, thereby lowering the electrical and thermal conductivities. As ion-irradiation damage is typically localised to within the first few microns of the sample, an accurate measurement of the change in thermal conductivity requires the use of carefully calibrated, surface-sensitive techniques, such as the {3-$\omega$} method \cite{cui2017thermal} or transient grating spectroscopy (TGS) \cite{hofmann2019transient}. In our simulations, we compute the thermal conductivity using the method developed by Mason \textit{et al.} \cite{mason2021estimate}, where atoms with a high potential energy are considered as scattering sites. We report thermal conductivities at room temperature.

Measurements of thermal conductivity in tungsten irradiated at higher temperature is sparse; data is available for {300\,K} \cite{cui2018thermal, Reza_AM_2020} and {1000\,K} \cite{cui2018thermal}.  We computed thermal conductivities for the {$10^{-4}$\,dpa/s} dataset, matching the dose rate of the experiments by Reza \textit{et al.} \cite{Reza_AM_2020}. Similar to the previous comparison, we find qualitative agreement between simulation and experiment at room temperature, with simulation overestimating the conductivity by about {30\,\%} at higher doses. A previous simulation study \cite{mason2021parameter} using the cascade annealing method found closer quantitative agreement. At high temperature, we find that simulation and experiment both show an increase in thermal conductivity, which can be attributed to a reduction in scattering sites as fewer crystal defects are present at high temperature. However, the simulations predict a near full recovery of thermal conductivity by {750\,K}, while the experiments indicate that recovery might not be quite as complete at {1000\,K}---however, the reported uncertainties are large, impeding a quantitative comparison.

To summarise, these comparisons suggest that the vacancy acceleration method overestimates the rate of vacancy and interstitial recombination compared to experiments in real materials. Furthermore, TEM characterisation of samples irradiated at comparable temperature and dose rate \cite{hu2021effect} suggest that the simulation method underestimates the number density of nanoscale-sized voids. A possible explanation for these two discrepancies might lie in the neglect of impurities in the simulated microstructure. Light element impurities, such as C, N, and O, can trap vacancies, and thereby suppress defect recombination whilst catalysing the void nucleation process \cite{song2024bridging}. Even the extra-high purity tungsten samples {(99.9999 wt.\%)} used by Hu \textit{et al.} \cite{hu2021effect} could contain up to {$\sim$\,10\,appm} of light element impurities, which would translate to {$\sim$\,10} impurity atoms in a 1 million atom system. In conclusion, while the effect of impurities on microstructural evolution is missing in the simulations performed here, it is feasible to investigate this in a future study.

\section*{Conclusion}
We developed a model that enables simulation of microstructural evolution under irradiation and simultaneous thermally driven recovery. This is achieved by accelerating specifically vacancy migration with a kinetic Monte Carlo algorithm, while the rest of microstructural evolution, including collision cascades and defect clustering, proceeds implicitly by standard molecular dynamics simulation. We discussed the rationale behind the methodology and outlined the algorithm. We studied the effect of the dose rate on the irradiation defect concentration in tungsten and presented a simple predictive rate theory model that matches the simulation result. We made three observations: First, the defect concentration is reduced as either the temperature increases or the dose rate decreases. Second, at higher temperatures, vacancies begin clustering into nanoscale voids. Third, the presence of a large dislocation loop in the initial microstructure leads to a higher defect concentration as it acts as a sink for primarily interstitial defects. These predictions are in qualitative agreement with experimental observations, however, an assessment of the quantitative accuracy is impeded by low volume of simulation data and sparsity of quantitative experimental data. The model appears to overestimate vacancy-defect recombination; a possible explanation is the present neglect of impurities in the model.

We conclude that a hybrid molecular dynamics/Monte Carlo method is a viable option for exploiting the gap in diffusion time-scales between interstitial-type defects and vacancies. The results of the model are consistent with the assertion that vacancies are the primary drivers of defect recombination, rather than interstitials. The method allows for an explicit and calibration-free description of sinks, as demonstrated here by inclusion of a large dislocation loop, and the simulation of individual vacancy migration events allows for the inclusion of impurity effects in a future study. However, we note that the simulation method becomes computationally impractical at higher temperatures due to the exponentially increasing number of vacancy migration events that need to be simulated in a given period of time.


\section*{Conflicts of interest}
There are no conflicts to declare.

\section*{Data availability}
The simulation script, atomistic simulation snapshots, and the data constituting the figures will be made openly available on \textsc{Zenodo} upon acceptance of the manuscript.

\section*{Acknowledgments}
We acknowledge useful discussions with Felix Hofmann, Thomas Jourdan, and Sabina Markelj. This work has been carried out within the framework of the EUROfusion Consortium, funded by the European Union via the Euratom Research and Training Programme (Grant Agreement No 101052200 -- EUROfusion) and from the EPSRC [grant number EP/W006839/1]. To obtain further information on the data and models underlying this paper please contact PublicationsManager@ukaea.uk. Views and opinions expressed are however those of the authors only and do not necessarily reflect those of the European Union or the European Commission. Neither the European Union nor the European Commission can be held responsible for them. The authors acknowledge the use of the Cambridge Service for Data Driven Discovery (CSD3) and associated support services provided by the University of Cambridge Research Computing Services (\url{www.csd3.cam.ac.uk}) in the completion of this work.

\appendix

\counterwithin{figure}{section}

\section*{Appendix}

\section{Vacancy detection validation}\label{app:vacdetection}

\begin{figure}[t]
\includegraphics[width=0.9\columnwidth]{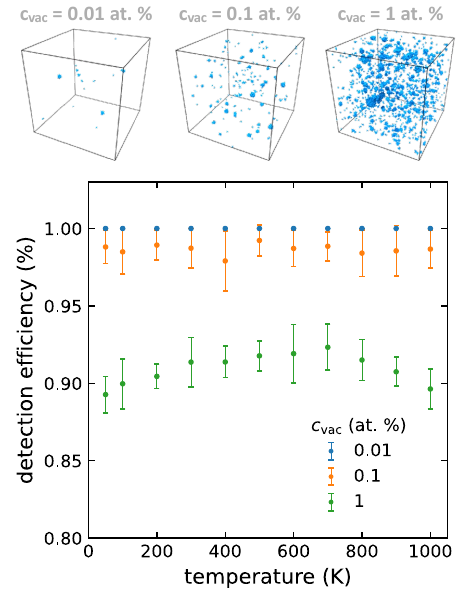}
\caption{Vacancy detection validation. Top: Microstructures with varying void volume fractions are analysed for monovacancies and divacancies as a function of temperature. Bottom: The fraction of detected over nominal vacancies, the  detection efficiency, is found to be insensitive to thermal noise.
}
\label{fig:A1}
\end{figure}

Here, we test the specificity and accuracy of the vacancy detection algorithm. We apply the algorithm to a tungsten simulation cell containing a void volume fraction of {0.01\,\%}, {0.1\,\%}, and {1\,\%}. The void structures are generated by iteratively deleting atoms in randomly placed spheres with radii $R$ drawn from the exponential distribution $R \sim \mathrm{Exp}(\lambda)$, here $\lambda = 1$ until the target void volume fraction is reached. The structures are analysed for monovacancies and divacancies using Wigner-Seitz Analysis, representing the ground truth data. Example void structures are shown in Fig.~\ref{fig:A1}.

Next, to introduce thermal noise consistent with a system temperature $T$, each atomic coordinate is perturbed by a random displacement vector $\vec{u}$, where each component is drawn from a normal distribution with zero mean and variance of  
\begin{equation}
\left\langle u_i \right\rangle^2 = a T + b T^2,
\end{equation}
where $a = 6.638\times 10^{-6}\textup{~\AA}^2\mathrm{K}^{-1}$ and $b = 2.710\times 10^{-9}\textup{~\AA}^2 \mathrm{K}^{-2}$. Parameters $a$ and $b$ were fitted to reproduce the mean-squared atomic displacement of BCC tungsten atoms at finite temperature but can also, in the linear approximation, be estimated from the Debye temperature $T_\mathrm{D} = 400\,\text{K}$ \cite{Kittel} via the Debye-Waller factor \cite{paradezhenko2017lattice}: 
\begin{equation}
\left\langle u_i \right\rangle^2 
= \frac{3\hbar^2}{k_\mathrm{B} M}\frac{T}{T^2_\mathrm{D}}
= 8.249\times 10^{-6}\frac{\textup{~\AA}^2}{\mathrm{K}} T
\end{equation}
where $M = 183.84\,\mathrm{u}$ is the atomic mass of tungsten.

Next, we run the vacancy detection algorithm on the perturbed structures, and test if the detected vacancies are consistent with the reference vacancies. We perform this analysis for temperatures spanning {50\,K} to {1000\,K}, for 10 independent perturbed structures per temperature, with results shown in Fig.~\ref{fig:A1}. We find that the vacancy detection algorithm is unaffected by thermal noise for the temperatures tested here. The vacancy detection algorithm becomes less accurate in detecting monovacancies and divacancies as the void volume fraction increases.

Further, we find that no vacancies were misclassified; there were no false positive detections of vacancies, and all of the identified vacancy clusters were identified correctly compared to the Wigner-Seitz ground truth. We note that at the highest tested void volume fraction, about {10\,\%} of the vacancies are not detected, suggesting that vacancies might be diffusing about {10\,\%} slower than they should. Solving Arrhenius' rate for the effective temperature, this slow-down effectively corresponds to a reduction in the simulated temperature on the order of {5\,K}. This error is comparatively small compared to the error due to the neglect of anharmonic effects in the vacancy hopping rate, see Appendix~\ref{app:kmc}, and hence we shall not discuss it further.

\section{Vacancy acceleration validation}\label{app:kmc}

The rkMC algorithm is validated by simulating the diffusion of a monovacancy without external stress, and with an external shear stress of $\sigma_\mathrm{xy} = 3\,\mathrm{GPa}$. Using standard MD, we sample the diffusion tensor over temperatures spanning {1000\,K} to {1700\,K}. Using the vacancy-accelerated method, we sample temperatures spanning {300\,K} to {900\,K}. To reconstruct the diffusion tensor in the MD simulation, we track the vacancy hops using an on-the-fly Wigner-Seitz Analysis method. In the accelerated simulation, we track the successful rkMC hopping events. 

\begin{figure}[t]
\centering
\includegraphics[width=0.75\columnwidth]{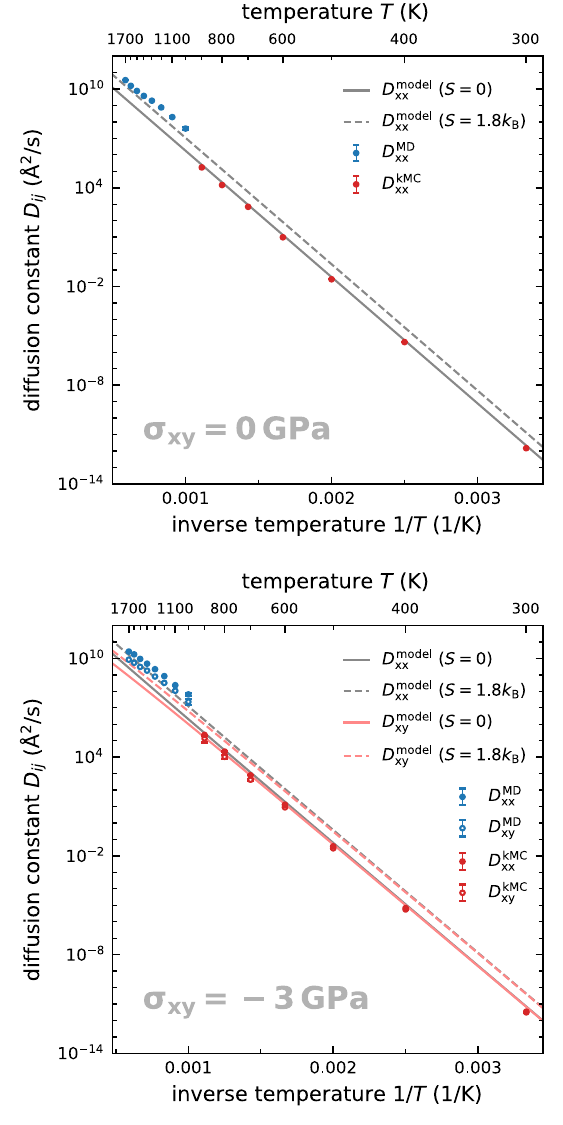}
\caption{Kinetic Monte Carlo validation. Shown are the diffusion tensors of a tungsten monovacancy, with and without external stress, obtained from standard molecular dynamics ($\tensor{D}^\mathrm{MD}$), from the vacancy acceleration method ($\tensor{D}^\mathrm{kMC}$), and from the analytical model ($\tensor{D}^\mathrm{model}$). The molecular dynamics and accelerated method results are consistent with the model predictions with ($S= 1.8k_\mathrm{B}$) and without ($S=0$) the entropic contribution to the migration barrier, respectively. Error bars indicating the standard error are too small to be visible on this axis.
}
\label{fig:A2}
\end{figure}

The simulated diffusion tensor is also compared to the analytical diffusion tensor. For a monovacancy at position $\vec{r}_0$ in a BCC crystal, it is given by
\begin{equation}
\tensor{D} = \frac{1}{6}\sum_{h=1}^{8}\left(\vec{d}_h\otimes \vec{d}_h\right) \nu_h^{\vec{d}},
\end{equation}
where the summation runs over the eight unique $\frac{1}{2}\left\langle 111 \right\rangle$ adjacent hopping sites placed at $\vec{r}_0 + \vec{d}_h$, and $\nu_h^{\vec{d}}$ is the stress-dependent hopping rate, see Eq.~\eqref{eq:hoppingrate}. For the comparison of the analytical model with molecular dynamics, we note that expression Eq.~\eqref{eq:hoppingrate} does not account for vibrational entropy, which, in the harmonic approximation, acts to effectively lower the migration barrier ${E_\mathrm{m}' = E_\mathrm{m} - T S}$ \cite{Vineyard_JPCS_1957}. For vacancy migration with this interatomic potential \cite{mason2017empirical}, we find $S = 1.8 k_\mathrm{B}$. We evaluate the model diffusion tensor including and excluding the entropic contribution.

Because we are validating the analytical model and accelerated simulations in reference to explicit molecular dynamics, we use the vacancy migration barrier and relaxation volume tensors consistent with the chosen interatomic potential \cite{mason2017empirical}. We determined the migration barrier using the nudged elastic band method \cite{henkelman2000climbing}, and the relaxation volume tensors following the method outlined in Ref.~\cite{MaPRM2019a}: $E_\mathrm{m} = 1.523\,\mathrm{eV}$, $\Omega_\mathrm{iso}^{(\mathrm{eq})} = -0.1221$, $\Omega_\mathrm{iso}^{(\mathrm{sd})} = -0.1897$, and $\Omega_\mathrm{aniso}^{(\mathrm{sd})} = -0.0800$ in atomic volumes.

The simulation results are shown in Fig.~\ref{fig:A2}. The results from the molecular dynamics and accelerated methods are in close agreement to the model prediction with and without the entropic contribution, respectively. In the absence of external shear stress, the dipole tensor is isotropic. When a shear stress of $\sigma_\mathrm{xy} = -3\,\mathrm{GPa}$ is included, the diffusion tensor gains off-diagonal elements $D_\mathrm{xy} = D_\mathrm{yz}$ with magnitude comparable to the diagonal elements, demonstrating that elastic stress is accounted for in the vacancy acceleration method.

\bibliography{reference}

\end{document}